\documentclass[aps,prl,twocolumn,floats]{revtex4}

\usepackage{amssymb}
\usepackage{amsmath}
\usepackage{graphicx}

\begin{document}

\title{Spin wave contribution to the nuclear spin-lattice relaxation
in triplet superconductors}

\author{Timofey Rostunov$^1$, Antoine  Georges$^2$, and Eugene Demler$^1$}
\affiliation{$^1$Department of Physics, Harvard University, Cambridge
MA 02138\\
$^2$Centre de Physique Theorique,
Ecole Polytechnique, 91128 Palaiseau Cedex, France}

\date{\today}

\begin{abstract}

We discuss collective spin wave excitations in triplet superconductors
with an easy axis anisotropy for the order parameter. Using a
microscopic model for interacting electrons we estimate the frequency
of such excitations in Bechgaard salts and ruthenate superconductors
to be one and twenty GHz respectively. We introduce an effective
bosonic model to describe spin-wave excitations and calculate their
contribution to the nuclear spin lattice relaxation rate. We find that
in the experimentally relevant regime of temperatures, this mechanism
leads to the power law scaling of $1/T_1$ with temperature. For two
and three dimensional systems the scaling exponents are three and five
respectively.  We discuss experimental manifestations of the spin wave
mechanism of the nuclear spin lattice relaxation.

\end{abstract}

\maketitle

Nuclear magnetic resonance (NMR) is a powerful tool for
analyzing ordered electron states in solids. NMR analysis has been
successfully applied to study magnetic insulators (see {\it e.g.}
\cite{beeman68} and references therein) as well as
several classes of unconventional superconductors, including high Tc
cuprates \cite{X1}, heavy fermion materials \cite{tou03}, ruthenates
\cite{X2}, and organic superconductors \cite{lefebvre00}.
In particular, NMR experiments have been useful for analyzing
the symmetry of the SC order parameter~\cite{lee02}
and for clarifying the structure of the phase diagram
in systems with competing orders~\cite{lee05}.

A common feature of the NMR experiments in certain families of triplet
superconductors (TSC) is the power law temperature dependence of the
nuclear spin lattice relaxation rate. Bechgaard
salts \cite{lee2000}, ruthenates
\cite{X2}, and heavy fermion materials \cite{X4} showed $1/T_1 \sim T^3$
at low temperatures and for small magnetic fields. Such behavior has
usually been interpreted as a signature of nodes in the quasiparticle
gap on the Fermi surface. Indeed, point and line nodes should lead to
$T^5$ and $T^3$ scaling of $1/T_1$ respectively. In several cases,
however, we have reasons to doubt the presence of nodes in the TSC
order parameter. For example, in quasi one dimensional Bechgaard salts
the natural order parameter has different signs on the two sheets of
the Fermi surface and no nodal points \cite{X5}. Also in ruthenates,
the order parameter that is consistent with spontaneous time reversal
breaking \cite{X6} and the quasi two dimensional nature of these
materials corresponds to a constant quasiparticle gap on the entire
Fermi surface \cite{Mackenzie}.  In this paper we consider a mechanism
of the nuclear spin lattice relaxation that is not due to Bogoliubov
quasiparticles but due to collective spin wave (SW) excitations of the
TSC order parameter. We demonstrate that in the experimentally
relevant regime of temperatures, this mechanism also leads to the
power law scaling of $1/T_1$ with temperature.

%%%%%%%%%%%%%%%%%%%%%%%%%%%%%%%%%%%%%%%%%%%%%%%%%%%%%%%%%%%%%%%%%%%%

Our starting point is the Moriya relation \cite{moriya63} for the
nuclear spin lattice relaxation rate
\begin{eqnarray}
\frac{1}{T\,T_1}= \frac{2\; g^2_\mathrm{N}\, |A|^2}{(g_\mathrm{eff}\mu_B)^2}
\int d^d\,q\,\,\,
\,\,\frac{\chi_{\perp H}''(q,\omega_N)}{\omega_N}
\label{moriya}
\end{eqnarray}
Here $A$ describes the strength of hyperfine interactions between
nuclear spins and conduction electrons, $g_N$ is a gyromagnetic ratio
of the nucleus, $g_\mathrm{eff}$ is an effective gyromagnetic ratio of
conducting electrons, $\mu_B$ is a Bohr magneton, $\chi_{\perp
H}''(q,\omega_N)$ is the imaginary part of the transverse ({\it i.e.}
perpendicular to the magnetic field) electron spin susceptibility
taken at the nuclear Larmor frequency $\omega_N$.
In the case of a perfect spin SU(2) symmetry, linearly dispersing SW
excitations exist down to arbitrarily small energies. In real
materials there is always  spin anisotropy which gives rise to
a finite gap for spin excitations, $\omega_0$. Below, we estimate the
value of $\omega_0$ to be tens of millikelvin for Bechgaard salts and
hundreds of millikelvin for the ruthenates. This is
much larger than the nuclear Larmor frequency, $\omega_N$, but smaller
than the typical temperature used in experiments.
When $\omega_0$ is much larger than $\omega_N$, creation
and annihilation of individual SWs does not affect $\chi''(\omega_N)$.
However, there is a contribution due to the scattering of
thermally excited SW excitations.
Let $\rho(E)$ be the density of states for SW excitations and
$n(E)=(\exp(E/k_BT)-1)^{-1}$ be the Bose distribution function.  From
the second order perturbation theory we have $ \chi''_{zz}(\omega_N)
\sim \int \rho(E) \rho(E+\omega_N) [ n(E)-n(E+\omega_N)] dE $.  The
characteristic energy scale in this integral is set by the temperature
$T$. Since $T \gg \omega_0$ we can assume linear dispersion of SW
excitations and take $\rho(E) \sim E^{d-1}$, where $d$ is the number
of spatial dimensions. Using $\omega_N \ll T$, we have
$\chi''_{zz}(\omega_N) \sim \omega_N \int E^{2d-2} (-\partial n
/\partial E) dE \sim T^{2d-2}$.  Combining this result with the Moriya
relation (\ref{moriya}) we obtain $1/T_1 \sim T^{2d-1}$. This simple
analysis does not take into account coherence factors in the
expression for $\chi''$.  Below we demonstrate that coherence
factors do not modify the scaling exponent of the nuclear spin lattice
relaxation rate in TSC. This is
in contrast to antiferromagnetic systems that also have linearly
spin waves, but in which coherence factors contribute an additional
$1/T^2$ factor to $1/T_1$ \cite{beeman68}.

For a detailed analysis of the nuclear spin lattice
relaxation rate we introduce an effective model that captures the
essence of triplet superconductivity and allows us to analyze collective
excitations. A simple picture of the TSC state corresponds to binding
electrons into Cooper pairs with spin one and momentum zero and Bose
condensing such pairs. In an effective bosonic model one can neglect
details of the orbital nature of Cooper pairs and consider them as
``elementary'' particles. Interactions are important for the correct
description of collective excitations, thus we are led to the Hubbard
type model for spin one bosons \cite{demler02}
\begin{eqnarray}
{\cal H}&=& -t \sum_{\langle ij \rangle\sigma}
(a^\dagger_{i\sigma} a_{j\sigma} +  a^\dagger_{j\sigma} a_{i\sigma})
 - \delta r \sum_i a_{i0}^\dagger a_{i0}
\nonumber\\
&+&\frac{U_0}{2} \sum_i n_i (n_i-1)
+\frac{U_2}{2} \sum_i \vec{S}_i^2-\mu \sum_i n_i
\label{Hubbard}
\end{eqnarray}
Here $a^\dagger_{i\sigma}$ creates a boson on site $i$ with spin
$\sigma=\{-1,0,1\}$. Operators $n_i$ and $\vec{S}_i$ describe the
number of atoms and the total spin on site $i$: $ n_i = \sum_\sigma
a^\dagger_{i\sigma} a_{i\sigma} $, $ \vec{S}_i = \sum_{\sigma,\sigma'}
a^\dagger_{i\sigma} \vec{T}_{\sigma \sigma'} a_{i\sigma'} $, where
$\vec{T}_{\sigma \sigma'}$ are the usual spin matrices for spin one
particles. The first term in the Hamiltonian (\ref{Hubbard}) describes
tunneling of Cooper pairs between neighboring lattice sites $i$ and
$j$.  An important aspect of the model (\ref{Hubbard}) is the presence
of two types of interaction terms.  The third term in (\ref{Hubbard})
depends only on the number of particles on each site and is the same
as for spinless bosons.  The fourth term in (\ref{Hubbard}) gives spin
dependence to the interaction (without breaking the spin SU(2)
symmetry) and is a novel feature of spinful Cooper pairs. The sign of
$U_2$ determines the difference between unitary ($U_2>0$) and
nonunitary ($U_2<0$) triplet superconductors. It is generally believed
that triplet pairing between fermions leads to unitary Cooper pairs
\cite{Mackenzie}.  Thus, from now on we assume that $U_2$ is
positive.  For concreteness, we consider a $d$ dimensional ($d=2,3$)
hypercubic lattice. Our results, however, do not depend on the precise
lattice structure.  The second term in equation (\ref{Hubbard})
introduces easy axis anisotropy for the order parameter by making the
condensation of $a_0$ to be energetically favorable. A state with
finite $\langle a_0 \rangle$ corresponds to the unitary state of
Cooper pairs with the $\mathbf{d}$ vector pointing along the $z$ axis
(see eq. (\ref{action1}) for the definition of the $\mathbf{d}$
vector).

 In the mean field approximation, we take
$
\langle a_{0} \rangle = \Psi_0
$
with
$
| \Psi_0 | ^2={(\mu+zt+\delta r)}/{U_0}
$
and $z=2d$ being the coordination number.
Without loss of generality, we can take $\Psi_0$ to be real.
Fluctuations in the phase
of $a_0$ correspond to the density (Bogoliubov) mode.
To find SW excitations we need to
consider $a_{\pm}$ operators. In the Hamiltonian (\ref{Hubbard}) we
replace $a_0$ by its expectation value, take the terms quadratic in
$a_{\pm}$, and perform the Bogoliubov rotation $
a_{i\alpha}=\frac{1}{\sqrt{N}}\sum_{\vec{k}} a_{\vec{k}\alpha}
e^{i\vec{k}\vec{r}_i} $,
$
a_{\vec{k}+}= v_k \gamma_{\vec{k}+}+u_k \gamma^\dagger_{-\vec{k}-},
a_{-\vec{k}-}= u_k \gamma^\dagger_{\vec{k}+}+ v_k \gamma_{-\vec{k}-}
$.
  Here $N$ is the
total number of lattice sites, $ u_k^2+v_k^2={(\xi_k+\Delta)}/{E_k} $,
$ 2u_kv_k=-{\Delta}/{E_k} $, $
\xi_k=-2t \sum_{n=1}^d \cos (k_n b) + U_0 \Psi_0^2 -\mu
$.
In these equations  $b$ is the lattice constant and $\Delta=U_2 \Psi_0^2$.
We obtain the diagonalized spin-wave Hamiltonian
$
{\cal H}_{0}= \sum_k E_k \left( \gamma^\dagger_{{k}+}\gamma_{{k}+}
+ \gamma^\dagger_{{k}-}\gamma_{{k}-} \right)
$
with
$
E^2_k=\xi_k^2+2 \Delta \xi_k
$.
Operators
$\gamma^\dagger_{k\pm}$ create SW excitations with
$S_z=\pm 1$.  In the long wavelength
limit we find
$
E^2_k =\omega_0^2+v_s^2k^2
$
with $\omega_0^2=\delta r^2 + 2 \Delta \delta r $ and
$v_s^2={2t\,U_2\,\Psi_0^2 a^2}$.

\begin{figure}
\includegraphics{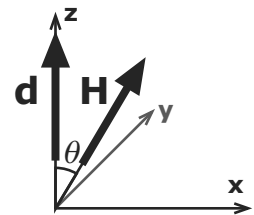}
\caption{Orientation of coordinate axes, $\mathbf{H}$, and the $\mathbf{d}$ vector.}
\label{fig.axes}
\end{figure}

We need to calculate the electron spin susceptibility in the direction
perpendicular to the applied magnetic field. Let $\theta$ be the angle
between the $z$ axis ({\it i.e.} the direction of the $\vec{d}$ vector)
and the direction of the
magnetic field (see Fig.~\ref{fig.axes}). We have
\begin{eqnarray}
\chi_{\perp H}=\sin^2\theta
\,\chi_{zz}+(1+\cos^2\theta)\,\chi_{xx}.
\label{chi_perp}
\end{eqnarray}
It is easy to see that
$\chi_{xx}$ and $\chi_{zz}$ are given by the correlation function of SWs
\begin{eqnarray}
\chi_{xx}(q,\omega)&=& 2 (g_{\rm eff} \mu_B \Psi_0)^2 (u_q+v_q)^2
\, \sum_{\alpha\beta} \, D^R_{\alpha\beta}(q,\omega)
\nonumber\\
\chi_{zz}(q,\omega)&=& (g_{\rm eff} \mu_B)^2
\, \sum_{\alpha\beta} \, \int \frac{d^d\!k\, d\Omega}{(2\pi)^{d+1}}\,U^2_{\alpha\beta}(k,k+q)\times
\nonumber\\
&&D^R_{\alpha\alpha}(k+q,\Omega+\omega)
D^A_{\beta\beta}(k,\Omega)
\label{Chi_Expression}
\end{eqnarray}
where we introduced the Nambu-Gorkov type notations
$
D^{R,A}_{\alpha\beta}(q,\omega)  = \int \,dt \, e^{i\omega t}
\theta(\pm t) \langle \mathrm{T}\, \Psi_{q\alpha}(t) \Psi_{q\beta}^\dagger(0) \rangle
$,
with
$
\Psi_{q\alpha} = \{ \gamma_{q+}, \gamma_{-q-}^\dagger \}^T,
$
and
$
U_{\alpha\beta}(k,k')=\delta_{\alpha\beta}(v_k v_{k'} - u_k u_{k'})
+ (1-\delta_{\alpha\beta})(u_k v_{k'} - v_k u_{k'}).
$

\begin{figure}
\includegraphics[width=8cm]{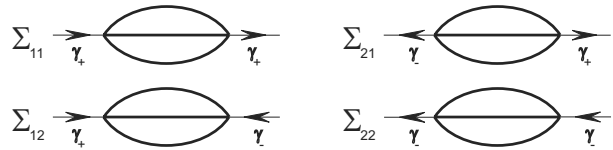}
\caption{Second order self-energy diagrams that gives rise to
scattering of spin waves.
For details see \cite{unpublished}.}
\label{fig.sigma2}
\end{figure}

Note that there is a qualitative difference in calculating $\chi_{zz}$
and $\chi_{xx}$. A non-uniform magnetic field in the $z$ direction can
scatter the existing thermally excited SWs. Thus we find finite
imaginary part of $\chi_{zz}$ at small frequencies by considering
the quadratic Bogoliubov Hamiltonian. We obtain
\begin{eqnarray}
&&\chi''_{zz}(\omega_N,q)=2(g_\mathrm{eff}\mu_B)^2
\frac{\omega_N}{T} \,b^d \,\,\int \frac{d^d k}{(2\pi)^d}\,
\nonumber\\
&\times&
 n(E_k)(n(E_{k+q})+1)\delta(E_k-E_{k+q}).
\label{eq.chi.zz}
\end{eqnarray}
By contrast, a non-uniform magnetic field in the $x$ direction can only
create or annihilate SW excitations. However, energies of these
excitations cannot be smaller than $\omega_0$. Hence, if we limit
ourselves to the quadratic Hamiltonian for SWs, we find that
$\chi''_{xx}$ is identically zero for frequencies smaller than
$\omega_0$ at any
temperature.  To get finite $\chi_{xx}''$ at small frequencies we need
to consider interactions between SWs.  Taking quartic terms in
equation (\ref{Hubbard}) and using definitions of SW operators,
we obtain the interaction terms between SW
excitations. These can be used to calculate self-energies for SW
excitations as shown in Fig~\ref{fig.sigma2}. We find
\cite{unpublished}
\begin{eqnarray}
&&\chi_{xx}''(q,\omega_N)=
(g_\mathrm{eff}\mu_B\Psi_0)^2\frac{\omega_N}{4\Delta T}
(U_0+3U_2)^2 \,b^{2d} \nonumber\\
&\times &\,\int \frac{d^d k_1}{(2\pi)^d}\,\int \frac{d^d k_2}{(2\pi)^d}\,
\frac{E_{k1}^2+E_{k2}^2+E_{k3}^2}{E_{k1}E_{k2}E_{k3}}
\nonumber\\
&\times&(n(E_{k1})+1)(n(E_{k2})+1)n(E_{k3})
\nonumber\\
&\times&\delta(E_{k1}+E_{k2}-E_{k3}).
\label{chi_final_experssion}
\end{eqnarray}
where $\vec{k}_3=\vec{k}_1+\vec{k}_2$.
We emphasize that equations (\ref{eq.chi.zz}) and
(\ref{chi_final_experssion}) apply only in the low
frequency limit $\omega_N \ll \omega_0$ which is relevant for
experiments.

Expressions (\ref{eq.chi.zz}) and (\ref{chi_final_experssion})
show that in equation (\ref{chi_perp})
contributions to $1/T_1$ due to $\chi''_{zz}$ and $\chi_{xx}''$ scale
as $T^{2d-1}$ and $T^{3d-2}$ respectively. For two and three
dimensional systems the $\chi''_{zz}$ contribution dominates (one can
check that this conclusion remains when we include prefactors). At
first glance this result appears surprising. Firstly, the real static
susceptibility is finite in the direction perpendicular to the
$\mathbf{d}$ vector but is zero along it.  Secondly, in the case of
full SO(3) symmetry, creation and annihilation of individual SWs
contribute to $\chi''_{xx}$ and $\chi''_{yy}$ at small frequencies,
but have no effect on $\chi''_{zz}$. A crucial part of our analysis is
the existence of a  spin gap $\omega_0$ which is much
larger than the nuclear Larmor frequency $\omega_N$. In this case
$\chi''(\omega_N)$
does not have contributions due to creation or annihilation of
individual SWs. For $\chi''_{zz}$ we take thermally
excited SWs and scatter them by the magnetic field. For
$\chi''_{xx}$ we also need thermally excited SWs but in
addition we must rely on interactions between them. SWs are
pseudo-Goldstone modes. At low energies interactions between them are
suppressed. This gives rise to the smallness of $\chi''_{xx}$ relative to
$\chi''_{zz}$.

To summarize, for two and three dimensional systems, we find that as
long as $\theta$ is not anomalously small, the SW contribution
to the nuclear spin lattice relaxation rate is given by
\begin{eqnarray}
\frac{1}{T_1} = \sin^2 \theta \,\,
|A|^2 g_{\rm N}^2\,
\frac{b^d}{v_s^{2d}}\,\frac{S_d^2}{4\pi^2}\;|B_{2d-2}|\;
T^{2d-1},
\label{T1_general_angle}
\end{eqnarray}
Here $S_d=2\pi^{d/2}/\Gamma(d/2)$ is the surface area of a unit sphere, and
$B_n$ are Bernoulli numbers, $B_2=1/6$, $B_4=-1/30$.
We remind the readers that equation (\ref{T1_general_angle})
applies when $T\gg\omega_0$.  At low
temperatures, $T\ll\omega_0$, we expect $1/T_1$ to start decreasing
exponentially, reflecting the exponential suppression
in the number of thermally excited SW excitations.
Equation (\ref{T1_general_angle}) also predicts that the nuclear
spin lattice relaxation rate should be very sensitive to the
direction of the magnetic field. We note, however,
that this argument is valid only for magnetic fields that are smaller
than the so-called spin-flop magnetic field, $H_\mathrm{flop}$.
In magnetic fields larger than $H_\mathrm{flop}$, the order
parameter $\vec{d}$ will always be perpendicular
to the direction of the applied field \cite{unpublished}.

It is useful to compare contributions to $1/T_1$ from magnons to the one
from quasiparticles. For concreteness we consider a quasi two
dimensional system with the TSC order parameter
$\vec{d}=\Delta_0\hat{z}k_x/k_F$, that has a line of nodes along the
$\hat{z}$ axis.  The quasiparticle contribution to the NMR relaxation
rate in such a state~\cite{sigrist91} is given by  $1/T_{1\,\mathrm{qp}}
=
\pi^2 |A|^2 g_\mathrm{N}^2 b^2 m^2 T^3 /6 \Delta_0^2$.
For comparison, we take equation
(\ref{T1_general_angle}) and use the BCS expressions for the velocity
of SW excitations (see below). We find $
{T_{1\,\mathrm{mag}}}/{T_{1\,\mathrm{qp}}}
=\pi^2\,{\Delta_0^2}/{\epsilon_F^2} $.  In a typical superconductor,
the value of the quasiparticle gap is much smaller than the electron
Fermi energy. Therefore, when we have both gapless quasiparticles and
SWs, the quasiparticle contribution will strongly dominate.
Only when the quasiparticles are fully gapped out do the magnons provide the
dominant contribution to $1/T_1$.

%%%%%%%%%%%%%%%%%%%%%%%%%%%%%%%%%%%%%%%%%%%%%%%%%%%%%%%%%%%%%%%%%%%%%%%

Now we outline the key steps of the analysis
that allowed us to estimate  the characteristic frequency of
SW excitations in TSC. We consider a phenomenological BCS type
model for interacting electrons
\begin{eqnarray}
{\cal H} &=& \sum_{k\sigma} \epsilon_k  c_{k\sigma}^\dagger c_{k\sigma}
- \sum_{a q} V_a^{-1}(q) d^\dagger_a(q) d_a(q)
\nonumber\\
d^\dagger_a(q) &=& \frac12 \sum_{k\alpha\beta}
V_a(q)\, f_k \,
c^\dagger_{k+q/2,\alpha} (i\sigma_2 \sigma_a)_{\alpha\beta}
c^\dagger_{-k+q/2,\beta}
\label{action1}
\end{eqnarray}
Here $d_a(q)$ are Fourier components of the TSC order parameter in the
direction $a$ ($a=x,y,z$), $f_k$ is an orbital wavefunction
(e.g. $f_k=\mathrm{sign}(k)$~\cite{X5}), and $c^\dagger_{k\sigma}$ are electron creation
operators. Parameters $V_a$ describe electron
interactions in the p-wave channel. A homogeneous unitary triplet
superconducting state is the ground state of the Hamiltonian
(\ref{action1}). Assuming easy axis anisotropy with $ V^z > V^x=V^y $,
we find $ \langle d_z(q=0) \rangle =\Delta_0 $ and $\langle d_x
\rangle = \langle d_y \rangle =0$.  To find SW excitations we
consider fluctuations of the TSC order parameter $ d_x(r,t) =
d^*_x(r,t) $. Integrating out fermions gives an effective action for
SW excitations \cite{unpublished} $ {\rm S}_{\rm eff} \{\,d_x
\,\} = \int_{\omega q} D_{x} (q,\omega) \,\,|\, d_x(q,\omega) \,|^2 $,
where $q$ and $\omega$ are the wavevector and the real frequency of
the SW.  For the spherically symmetric Fermi surface in $d$
dimensions, we have $
D_{x}(\omega,q)= \frac14 N_0 ( \omega_0^2 - \omega^2 + v_s^2 q^2 ) $.
Here $N_0$ is the density of states at the Fermi energy, $v_s^2 =
{v_F^2}/{d}$, $v_F$ is the Fermi velocity, and
\begin{eqnarray}
\omega_0^2 = 4 \Delta_0^2 N^{-1}_0({V^{-1}_x} - {V^{-1}_z} )
\label{Omega_zero}
\end{eqnarray}
Zeroes of the function $D_{x}$ correspond to SW excitations.  In the case of
spin symmetric interactions with $V_z=V_x$, SWs are Goldstone
modes of broken spin symmetry and are gapless.  They have linear
dispersion with the velocity $v_s$.  In the case of the easy axis
anisotropy, SWs have a gap $\omega_0$.

As a concrete example, we consider the TSC state in Bechgaard
salts. We assume that the
spin anisotropic part of the interaction in this phase is the same as
in the antiferromagnetic state of this material $
\Delta {\cal H}_{\rm anis} = \delta J_z \sum_{\langle ij \rangle}
S_i^z S_j^z
$.
Here $\langle ij \rangle$ corresponds to the nearest neighbor sites and
the spin $z$ axis points along the crystallographic $b'$ axis.
The antiferromagnetic resonance
experiments
\cite{torrance82}
suggest $\delta J_z = 0.01K$.  We can express
$
\Delta {\cal H}_{\rm anis}$ in the form similar to equation
(\ref{action1}) with $ - \delta V^z= \delta V^x = \delta V^y =
\frac{1}{2} \delta J_z v_0 $.  Here $v_0$ is the volume of the unit
cell.  Assuming that the anisotropic term is a
small correction to the spin symmetric interaction, the total
interaction entering equation (\ref{action1}) is $V^a= V+ \delta
V^a$.  The value of $V$ can be estimated from the BCS equation for
the transition temperature
$ T_c= 1.14\;\omega_{\rm BOS} e ^{-1/{N_0 V}} $.  Here $\omega_{\rm
BOS}$ is the characteristic frequency of bosons providing electron
pairing.  Combining all expressions, we find
\begin{eqnarray}
\frac{\omega_0}{\Delta_0}=2(\delta J_z\, N_0\, v_0)^{1/2}\;\log (\frac{1.14\,\omega_{BOS}}{T_c})
\end{eqnarray}
Taking $ \delta J_z=0.01K$, $N_0=2\cdot 10^{33}\;erg^{-1}cm^{-3}$,
$\omega_{\rm BOS}=1000$K, $T_c=1.4\,K$, $\Delta_0=2.5\,K$, and
$v_0=360\,\AA^3$ we obtain $\omega_0$ around one GHz.  There is a
simple qualitative argument that supports our result that $\omega_0$
for Bechgaard salts lies in the GHz range. This argument relies on a
comparison of  SW resonances in the antiferromagnetic and
superconducting states of these materials. In quasi one-dimensional
systems
microscopic descriptions of the
two states are
similar and BCS type models and our analysis
of SW excitations can be used in the antiferromagnetic phase as
well.  From the discussion above, we expect that the ratio of SW energies
in the two phases is approximately proportional to the
ratio
of AF and TSC
transition temperatures, i.e. around ten.
In the AF phase the SW resonance
frequency was measured in the tens of GHz range
\cite{torrance82}.
Hence,
in the superconducting state it should be a factor of ten smaller,
which brings us into the GHz range.

Similar analysis can be done for the TSC state in $Sr_2RuO_4$.
Sigrist and coworkers (see e.g. Ref \cite{X9}) showed that spin-orbit
coupling in these materials leads to a difference in the transition
temperature of various order parameters of the order of two percent.
From the BCS expression for $Tc$ we have $\delta T_c/T_c = \delta V/ N_0
V^2$.  Taking $\Delta_0=2.6$K, from equation (\ref{Omega_zero})
we estimate $\omega_0$ to be around
twenty GHz.

$Sr_2RuO_4$ is a quasi two dimensional material.
Bechgaard salts have a mixed dimensionality
\cite{schwartz98,lebed05} with easy, intermediate,
and hard directions in transport. Hence, they may also exhibit
properties of a quasi two dimensional system with respect to SW
excitations.
Both the ruthenates and Bechgaard salt superconductors
exhibit the $T^3$ dependence of $1/T_1$
at low temperature. Therefore these materials should be good
candidates for experimental investigation of the SW
mechanism of nuclear spin lattice relaxation discussed in this paper.

In summary, we studied spin wave excitations in triplet
superconductors with the easy axis spin anisotropy. We derived an
explicit expression for the energy of such excitations and used it to
estimate the spin wave energy in Bechgaard salts and ruthenate
superconductors.  We considered the effects of spin wave excitations
on the nuclear spin lattice relaxation rate. We showed that in the
experimentally relevant regime of temperatures, $1/T_1$ has a power
law scaling with temperature, including the $T^3$ dependence for two
dimensional systems. We showed that the spin wave mechanism predicts a
dramatic decrease of $1/T_1$ for temperatures lower then the
energy of spin wave excitations and leads to dependence of $1/T_1$
on the direction of the applied magnetic field.

We are grateful for useful discussions with E. Altman, D. Podolsky,
and A. Polkovnikov.  This work was supported by the NSF grant
DMR-0132874, the Sloan Foundation, and KITP UCSB.

 \end{document}